# Fokker-Planck and Chapman-Kolmogorov equations for Ito processes with finite memory

Joseph L. McCauley[+]
Physics Department
University of Houston
Houston, Tx. 77204-5005
jmccauley@uh.edu

[+]Senior Fellow
COBERA
Department of Economics
J.E.Cairnes Graduate School of Business and Public Policy
NUI Galway, Ireland

## Abstract

The usual derivation of the Fokker-Planck partial differential eqn. (pde) assumes the Chapman-Kolmogorov equation for a Markov process [1,2]. Starting instead with an Ito stochastic differential equation (sde) we argue that finitely many states of memory are allowed in Kolmogorov's two pdes, K1 (the backward time pde) and K2 (the Fokker-Planck pde), and show that a Chapman-Kolmogorov eqn. follows as well. We adapt Friedman's derivation [3] to emphasize that finite memory is not excluded. We then give an example of a Gaussian transition density with 1-state memory satisfying both K1, K2, and the Chapman-Kolmogorov eqns. We begin the paper by explaining the meaning of backward diffusion, and end by using our interpretation to produce a new, short proof that the Green function for the Black-Scholes pde describes a Martingale in the risk neutral discounted stock price.

## 1. The meaning of Kolmogorov's first pde

Consider a diffusive process described by an Ito stochastic differential equation (sde) [3,4] with or without finite memory in the drift and diffusion coefficients,

$$dx = R(x,t)dt + \sqrt{D(x,t)}dB(t), \qquad (1)$$

where B(t) is the Wiener process. By finite memory, we mean explicitly a history of a finite nr. k of earlier states $(x_k,t_k;…;x_1,t_1)$. This means that R and D may depend not only on the present state (x,t) but also on history $(x_k,t_k;…;x_1,t_1)$, so that the forward-time 2-point transition density $p_2(x,t+T:y,t)$ for the Ito process

$$x(t+T) = x(t) + \int_t^{t+T} R(x(s),s)ds + \int_t^{t+T} \sqrt{D(x(s),s)}dB(s) \qquad (2)$$

also depends on history $(x_k,t_k;…;x_1,t_1)$, where $t \geq t-T \geq t_k \geq … \geq t_1$. First, however, we derive the pde for the backward time 2-point transition density.

Consider a measurable, twice differentiable dynamical variable A(x,t). The sde for A is (by Ito's lemma [3,4,5])

$$dA = (\frac{\partial A}{\partial t} + R\frac{\partial A}{\partial x} + \frac{D}{2}\frac{\partial^2 A}{\partial x^2})dt + \sqrt{D}\frac{\partial A}{\partial x}dB \quad (3)$$

so that

$$A(x(t+T),t+T) = A(x(t),t) + \int_t^{t+T} (\frac{\partial A(x(s),s)}{\partial t} + R\frac{\partial A}{\partial x} + \frac{D}{2}\frac{\partial^2 A}{\partial x^2})ds + \int_t^{t+T} \sqrt{D(x(s),s)} \frac{\partial A(x(s),s)}{\partial x} dB(s)$$
. (4)

A martingale is defined by the conditional average $<A(x,t+T)>_c = A(x,t)$ [3,5] where a backward in time average is indicated, and where $<x(t+T)>_c = x(t) + \int <R>_c ds$. We want to obtain the generator for the backward time transition probability density $p^+(x,t:y,t+T)$. This can be done in either of two ways. First, with $T>0$ we have by definition that

$$A(x,t) = \int p^+(x,t:y,t+T)A(y,t+T)dy. \quad (5)$$

or $A(x,t) = U^+(t,t+T)A(t+T)$ where $U^+$ describes the evolution backward in time (no assumption is made here of an inverse for forward time diffusion). For analytic functions A we have

$$A(x,t) \approx A(x,t+T) + \int p^+(x,t:y,t+T)((y-x)\frac{\partial A(x,t+T)}{\partial x} + (y-x)^2 \frac{\partial^2 A}{\partial x^2})dy + ...$$
(6)

so that with T vanishing, and using the usual definition of drift and diffusion coefficients [1], we obtain the backward time diffusion pde [4]

$$0 = \frac{\partial A(x,t)}{\partial t} + R(x,t)\frac{\partial A(x,t)}{\partial x} + \frac{D(x,t)}{2}\frac{\partial^2 A(x,t)}{\partial x^2} \quad (7)$$

if the moments $\int(y-x)^n p^+ dy$ vanish fast enough with T for $n \geq 3$. If the transition density depends on a *finite* history of exactly k earlier states, $p^+(x,t:y,s) = p^+_n(x,t:y,s;x_k,t_k;...;x_1,t_1)$ with k=n-2, then that history appears in the drift and diffusion coefficients as well, e.g.,

$$D(x,t;x_k,t_k,...,x_1,t_1) \approx \frac{1}{T}\int(y-x)^2 p_{k+2}^+(y,t:x,t-T;x_k,t_k;...;x_1,t_1)dy$$
(8)

as T vanishes.

Second, note that the backward time pde (7) follows directly from (4) simply by setting the drift term equal to zero, yielding a martingale

$$A(x(t+T),t+T) = A(x(t),t) + \int_t^{t+T}\sqrt{D(x(s),s)}\frac{\partial A(x(s),s)}{\partial x}dB(s)$$
(9)

In this second and more general derivation no assumption is made or needed either of the moments $\int(y-x)^n p^+ dy$ vanishing fast enough with T for $n \geq 3$, or of analyticity of $A(x,t)$.

We've made no assumption that A is positive. I.e., A is generally not a 1-point probability density, $A(x,t)$ is simply any martingale. By (5) the required transition density is the Green function of (7),

$$0 = \frac{\partial g^+(x,t:y,s)}{\partial t} + R(x,t)\frac{\partial g^+(x,t:y,s)}{\partial x} + \frac{D(x,t)}{2}\frac{\partial^2 g^+(x,t:y,s)}{\partial x^2}$$
(10)

where $g^+(x,t:y,t)=\delta(x-y)$. I.e., $p^+(x,t:y,s)=g^+(x,t:y,s)$ with t<s. The conditions under which $g^+$ exists, is unique and nonnegative definite are stated in Friedman [4]. Eqn. (10) is called Kolmogorov's first pde (K1) [1].

What does K1 mean? Simply that martingales can be constructed via Ito's lemma.

## 2. The Fokker-Planck pde with finite memory

Consider next *any* measurable twice-differentiable dynamical variable A(x(t)). A(x) is *not* assumed to be a martingale. The time evolution of A is given by Ito's lemma [6,7]

$$dA = (R\frac{\partial A}{\partial x} + \frac{D}{2}\frac{\partial^2 A}{\partial x^2})dt + \sqrt{D}\frac{\partial A}{\partial x}dB. \quad (11)$$

We can calculate the conditional average of A, conditioned on $x_o$ at time $t_o$ in x(t)=$x_o$+∫R(x,s)ds+∫√D(x,s)dB(s), forward in time if we know the transition density $p_2$(x,t:$x_o$,$t_o$)) forward in time,

$$\langle A(x(t))\rangle = \int p_2(x,t:x_o,t_o)A(x)dx. \quad (12)$$

Note that this is not the rule for the time evolution of a 1-point probability density. From

$$\frac{d\langle A(x(t))\rangle}{dt} = \int \frac{\partial p_2(x,t:x_o,t_o)}{\partial t} A(x)dx \quad (13)$$

and using

$$\langle dA\rangle = \left(\left\langle R\frac{\partial A}{\partial x}\right\rangle + \left\langle \frac{D}{2}\frac{\partial^2 A}{\partial x^2}\right\rangle\right)dt \quad (14)$$

with <dA>/dt defined by (13), we obtain from (14), after integrating twice by parts and assuming that the boundary terms vanish,

$$\int dx A(x)\left[\frac{\partial p_2}{\partial t} + \frac{\partial (Rp_2)}{\partial x} - \frac{1}{2}\frac{\partial^2 (Dp_2)}{\partial x^2}\right] = 0, \quad (15)$$

so that the transition density is the Green function of the Fokker-Planck pde [1-4], or Kolmogorov's second pde (K2)

$$\frac{\partial p_2}{\partial t} = -\frac{\partial (Rp_2)}{\partial x} + \frac{1}{2}\frac{\partial^2 (Dp_2)}{\partial x^2}. \quad (16)$$

Since $p_2$ is a transition density we also have the 2-point density $f_2(x,t;y,s)=p_2(x,t:y,s)p_1(y,s)$ where the 1-point density $f_1=p_1$ satisfies

$$p_1(x,t) = \int f_2(x,t;y,s)dy = \int p_2(x,t:y,s)p_1(y,s)dy \quad (17)$$

and so satisfies the same pde (16) as does $p_2$ but with an arbitrary initial condition $p_1(x,t_1)=f(x)$. Note the difference with (12). So far, no Markovian assumption was made.

In particular, no assumption was made that R, D, and hence $p_2$, are independent of memory of an initial state, or of finitely many earlier states. If there is memory, e.g. if $p_1(x,t_o)=u(x)$ and if $D=D(x,t;x_o,t_o)$ depends on one initial state $x_o=\int xu(x)dx$, then due to memory in $p_1(x,t)$ [8],

$$p_2(x_3,t_3:x_2,t_2) = \frac{\int p_3(x_3,t_3:x_2,t_2;x_1,t_1)p_2(x_2,t_2:x_1,t_1)p_1(x_1,t_1)dx_1}{\int p_2(x_2,t_2:x_1,t_1)p_1(x_1,t_1)dx_1}$$
, (18)

then by the 2-point transition density we must understand that $p_2(x,t:y,s)=p_3(x,t:y,s;x_o,t_1)$. That is, in the simplest case $p_3$ is required to describe the stochastic process. Memory appears in (18) if, e.g., at time $t_o$ $u(x)=\delta(x-x_o)$ with $x_o\neq 0$ [8]. The main idea is that we are dealing quite generally with Ito sdes and corresponding pdes for transition densities with memory of a finite nr. n-2 of states, so that the 2-point transition density is $p(x,t:y,s)=p_n(x,t:y,s;x_{n-2},t_{n-2};\ldots;x_1,t_1)$ depending on n-2 earlier states.

Now, for the case where A(x(t)) *is* a martingale, then (12) must yield

$$\langle A \rangle_t = \int p(x,t:x_o,t_o)A(x)dx = A(x_o), \qquad (19)$$

and since (19) cannot differ from (5) if the theory is to make any sense, then there must be a connection between the backward and forward time transition densities $p^+$ and $p_2$. Comparing (19) with (5) we see that $p^+$ and $p_2$ must be adjoints. For a Markov process it's very easy to use the Chapman-Kolmogorov eqn. to derive both [1] the Fokker-Planck and Kolmogorov's backward time pde, and then prove that the Green function of K1 is the adjoint of the Green function of K2, but we will avoid making any Markovian assumption in order to permit finite memory in the formalism. In particular, we have not and will not assume in advance that a Chapman-Kolmogorov eqn. holds, but will next explain why that eqn. follows, even with finite memory. Then, in part 4, we'll derive the Chapman-Kolmogorov eqn. (21) below from memory dependent pdes K1 and K2.

**3. The Chapman-Kolmogorov eqn. for finite memory**

That a Chapman-Kolmogorov eqn. should hold for finitely many states of memory follows from standard definitions of conditional probability densities. With an unstated, even infinite, number of states in memory the history-dependent 2-point transition densities obey the hierarchy

$$p_{k-1}(x_k,t_k|x_{k-2},t_{k-2};...;x_1,t_1) = \int dx_{k-1} p_k(x_k,t_k|x_{k-1},t_{k-1};...;x_1,t_1) p_{k-1}(x_{k-1},t_{k-1}|x_{k-2},t_{k-2};...;x_1,t_1)$$

. (20)

For fractional Brownian motion (fBm), e.g., there is no reason to expect this hierarchy to truncate. But consider processes where the memory is finite and of number n-2, so that $p_k=p_n$ for all k≥n. Then from (20) we obtain the Chapman-Kolmogorov eqn. in the form [9]

$$p_n(x_n,t_n|x_{n-1},t_{n-1};...;x_1,t_1) = \int dy\, p_n(x_n,t_n|y,s;x_{n-2},t_{n-2};...;x_1,t_1) p_n(y,s|x_{n-1},t_{n-1};...;x_1,t_1)$$
(21)

for a process with finite memory. Next, for completeness, we will take a step backward and show that the pde K1 for an Ito process (1) with finite memory in R and/or D implies both the Fokker-Planck pde and the Chapman-Kolmogorov eqn.

## 4. Ito implies K1 and K2 implies Chapman-Kolmogorov, even with finite memory

Consider the linear operators

$$L^+ = \partial/\partial t + R(x,t)\partial/\partial x + (D(x,t)/2)\partial^2/\partial x^2 \qquad (22)$$

and

$$Lu = -\partial u/\partial t + \partial(R(x,t)u)/\partial x - \partial^2(D(x,t)u/2)/\partial x^2, \qquad (23)$$

acting on a function space of measurable, twice (not necessarily continuously) differentiable functions satisfying boundary conditions at t=∞, and at x=-∞ and x=∞ to be indicated below. Both operators followed superficially independently above from the Ito process (1), but we can start with (22) and then obtain (23) via

$$uL^+v - vLu = \frac{\partial}{\partial t}(uv) + \frac{\partial}{\partial x}(vRu + \frac{1}{2}uD\frac{\partial v}{\partial x} - v\frac{1}{2}\frac{\partial uD}{\partial x}), \quad (24)$$

which is a form of Green's identity (see also [4], but where the operator L is studied in standard elliptic rather than in Fokker-Planck form). With suitable boundary conditions on u,v [4] then L and $L^+$ are adjoints of each other:

$$\int_0^\infty dt \int_{-\infty}^\infty (vLu - uL^+v)dx = 0. \quad (25)$$

Starting with an Ito process (1) and K1, we have deduced K2. No Markovian assumption has been made. Again, the formal conditions under which (25) holds are stated in Friedman [4].

Next, let $g^+(x,t:\xi,\tau)$ denote the Green function of K1, $L^+g^+=0$, and let $g(x,t:\xi,\tau)$ denote the Green function of K2, Lg=0. Let $\tau<s<t$ and assume also that $\tau+\varepsilon<s<t-\varepsilon$, which avoids sitting on top of a delta function. Integrating (24) over y from -∞ to ∞ and over s from $\tau+\varepsilon$ to $t-\varepsilon$ with the choices $v(y,s)=g^+(y,s:x,t)$ and $u(y,s)=g(y,s:\xi,\tau)$, we obtain [4]

$$\int g(y,t-\varepsilon:\xi,\tau)g^+(y,t-\varepsilon:x,t)dy = \int g(y,\tau+\varepsilon:\xi,\tau)g^+(\tau+\varepsilon:x,t)dy . \quad (26)$$

With ε vanishing and using $g(y,\tau:\xi,\tau)=\delta(y-\xi)$, $g^+(y,t:x,t)=\delta(y-x)$, we obtain the adjoint condition for the Green functions

$$g(x,t:\xi,\tau) = g^+(\xi,\tau:x,t). \quad (27)$$

Next, apply the same argument but with times $\tau \leq t'' \leq t' \leq t$ to obtain (instead of (26))

$$\int g(y,t':\xi,\tau)g(x,t:y,t')dy = \int g(y,t'':\xi,\tau)g(x,t:y,t'')dy. \quad (28)$$

If we let t″ approach τ, then we obtain the Chapman-Kolmogorov eqn.

$$g(x,t:\xi,\tau) = \int g(x,t:y,t')g(y,t':\xi,\tau)dy, \quad (29)$$

again, *without having made any Markovian assumption*. The considerations of parts 2 and 3 tell us that we must restrict to transition densities depending at worst on only finitely many states in memory.

Summarizing, beginning with the Ito sde (1) and obtaining K1 (10) we've deduced K2 and finally the Chapman-Kolmogorov eqn. The derivation follows that of [4] where a Markov process was claimed, but we see that nowhere was the assumption of a Markov process either used or needed. The implication is that, with suitable boundary conditions on Green functions, an Ito sde implies both K1 and K2 and the Chapman-Kolmogorov eqn., even with finite memory (eqn. (21) may make no sense even for countably infinitely many states in memory, and demonstrably does not hold for non-Ito processes like fBm [7,10]).

To show that this new formalism is not vacuous, we now provide a simple example. We provide no example for variable diffusion D(x,t) where the (x,t) dependence is not separable, because even for the scaling class of models [6] we do not yet know how to calculate a model green function analytically.

## 5. A Gaussian process with 1-state memory

Consider first the 2-point transition density for an arbitrary Gaussian process in the form [8]

$$p(x,t:y,s) = \frac{1}{\sqrt{2\pi K(t,s)}} e^{-(x-m(t,s)y-g(t,s))^2/2K(t,s)}. \quad (30)$$

Until the pair correlation function $<x(t)x(s)>$ α $m(t,s)$ is specified, no particular process is indicated by (30). Processes as wildly different and unrelated as fBm [10], scaling Markov processes [10], and Ornstein-Uhlenbeck proceses [11] are allowed. Depending on the pair correlation function $<x(t)x(s)>$, memory, including long time memory, may or may not appear. To obtain fBm, e.g., g=0 and $<x(s)x(t)>$ must reflect the condition for stationary increments [10], which differs strongly from a condition of time translational invariance whereby m, g, and K may depend on (s,t) only in the form s-t. Fortunately, Hänggi and Thomas [8] have stated the conditions for a Gaussian process (30) to satisfy a Chapman-Kolmogorov eqn., namely,

$$m(t,t_1) = m(t,s)m(s,t_1)$$
$$g(t,t_1) = g(t,s) + m(t,s)g(s,t_1) \quad . \quad (31)$$
$$K(t,t_1) = K(t,s) + m^2(t,s)K(s,t_1)$$

Actually, Hänggi and Thomas stated in [8] that (31) is the condition for a *Markov* process, but we will show that the Chapman-Kolmogorov condition (31) is satisfied by at least one Gaussian process with memory.

Consider next the 1-point density $p_1(x,t)$ for a specific Ito process with simple memory in the drift coefficient, the Shimizu-Yamato model [9,12]

$$\frac{\partial p_1}{\partial t} = \frac{\partial}{\partial x}((\gamma+\kappa)x - \kappa\langle x\rangle + \frac{Q}{2}\frac{\partial}{\partial x})p_1 \qquad (32)$$

with initial data p(x,t$_o$)=u(x) and with <x>=∫xp$_1$(x,t)dx. The parameter Q is the diffusion constant. Since the drift coefficient in (1) is R=-(γ+κ)x+κ<x>, and since [13]

$$\frac{d\langle x\rangle}{dt} = \langle R\rangle = -\gamma\langle x\rangle \qquad (33)$$

we obtain

$$\langle x\rangle = x_o e^{-\gamma(t-t_o)} \qquad (34)$$

where

$$x_o = \int xu(x)dx. \qquad (35)$$

This provides us with a drift coefficient with initial state memory,

$$R(x,t;x_o,t_o) = -(\gamma+\kappa)x + \kappa x_o e^{-\gamma(t-t_o)}. \qquad (36)$$

Because γ≠0 the memory cannot be eliminated via a simple coordinate transformation z=x-<x>.

The Fokker-Planck pde for the transition density p$_2$(x,t:y,s;x$_o$,t$_o$) is

$$\frac{\partial p_2}{\partial t} = \frac{\partial}{\partial x}((\gamma+\kappa)x - \kappa x_o e^{-\gamma(t-t_o)} + \frac{Q}{2}\frac{\partial}{\partial x})p_2 \qquad (37)$$

with $p_2(x,t:y,t;x_o,t_o)=\delta(x-y)$. The solution is a Gaussian (30) with 1-state memory where

$$m(t,s) = e^{-(\gamma+\kappa)(t-s)}$$

$$K(t,s) = \frac{Q}{\gamma+\kappa}(1-e^{-2(\gamma+\kappa)(t-s)}) \qquad . \qquad (38)$$

$$g(t,s) = x_o(e^{-\gamma(t-t_o)} - e^{-(\gamma+\kappa)t+\gamma t_o+\kappa s})$$

*An easy calculation shows that the Chapman-Kolmogorov conditions (31) are satisfied with finite memory $(x_o,t_o)$. Furthermore, $p^+(y,s:x,t;x_o,t_o)=p_2(x,t:y,s;x_o,t_o)$ satisfies the backward time diffusion pde K1 in the variables (y,s),*

$$0 = \frac{\partial p^+}{\partial s} + R(y,s;x_o,t_o)\frac{\partial p^+}{\partial y} + \frac{Q}{2}\frac{\partial^2 p^+}{\partial y^2} \qquad (39)$$

with drift coefficient

$$R(y,s;x_o,t_o) = -(\gamma+\kappa)x + \kappa x_o e^{-\gamma(s-t_o)}. \qquad (40)$$

*This shows that backward time diffusion makes sense in the face of memory.* That memory simply yields $p^+(y,t_o:x_o,t_o;x_o,t_o)= \delta(y-x_o)$.

## 6. Black-Scholes from a different standpoint

Recapitulating, we understand the meaning of backward time diffusion qualitatively: we can construct martingales from an Ito process via Ito's lemma by setting the drift coefficient equal to zero, yielding K1 (see Steele [5] for simple but instructive martingales that can be constructed by solving (7) for various different initial and boundary conditions). This insight allows us to prove directly from the risk neutral hedge, the so-called delta hedge [13], that the Black-Scholes pde describes a martingale in the risk neutral discounted 'stock' price. We begin with the sde for the stock price p(t),

$$dp = \mu p dt + p\sqrt{d(p,t)}dB \qquad (41)$$

where μ is the unreliably known or estimated 'interest rate' on the stock. In the delta hedge strategy, w(p,t) is the option price and satisfies the Black-Scholes pde [13]

$$rpw(p,t) = \frac{\partial w(p,t)}{\partial t} + rp\frac{\partial w(p,t)}{\partial p} + \frac{p^2 d(p,t)}{2}\frac{\partial^2 w(p,t)}{\partial p^2} \qquad (42)$$

where r is the risk free interest rate (the interest rate on a bank deposit, money market fund, or CD). With $v = we^{r(t-T)}$, where T is the expiration time of a 'European' option,

$$0 = \frac{\partial v(p,t)}{\partial t} + rp\frac{\partial v(p,t)}{\partial p} + \frac{p^2 d(p,t)}{2}\frac{\partial^2 v(p,t)}{\partial p^2}. \qquad (43)$$

By (6), v defined by (43) is a martingale so that w(p,t) describes a martingale in the risk neutral discounted option price. That is, this model predicts a theoretically 'fair' option price, and corresponds to a stock price S(t) where the interest rate is r,

$$dS = rSdt + S\sqrt{d(S,t)}dB. \qquad (44)$$

See [14] for a longer proof that the Green function for the Black-Scholes pde (42) describes a martingale in the risk neutral discounted stock price. From our standpoint, the Black-Scholes pde is simply a standard equation of martingale construction for Ito processes.

We end with two remarks. First, Friedman [4] shows that the Chapman-Kolmogorov eqn. is not restricted to K1 and its adjoint the Fokker-Planck pde, but holds more generally for Green functions of pdes of the type

$$L^+v = \partial v/\partial t + c(x,t)v + R(x,t)\partial v/\partial x + (D(x,t)/2)\partial^2 v/\partial x^2 = 0 \quad (45)$$

and its adjoint. With $c=-R=rx$ and $D=x^2 d(x,t)$, where $x=p$ is I this case the stock price, we obtain the Black-Scholes pde (42). That the Green function for the Black-Scholes pde obeys the Chapman-Kolmogorov eqn. is surprising.

Initial value problems of (45), where u(x,T) is specified at a forward time T>t, are solved by a Martingale construction that results in the Feynman-Kac formula [4]. Defining M(s)=v(x,s)I(s), with dv(x,s) given by Ito's lemma and using (45) we obtain

$$dM = dvI + vdI = -c(x,s)v(x,s)ds + v(x,s)dI(s) + \sqrt{D(x(s),s)}\frac{\partial v}{\partial x}I(s)dB(s) \quad (46)$$

We obtain a martingale M(s)=v(x,s) with the choice

$$I(s) = e^{-\int_s^t c(x(q),q)dq}, \quad (47)$$

so that the solution of (45) is given by the martingale condition M(t)=<M(T)>,

$$v(x,t) = \left\langle v(x(T),T)e^{\int_t^T c(x(s),s)ds} \right\rangle \quad (48)$$

where the Feynman-Katz average (48) at time T is calculated using the Green function $g^+(x,t:x(T),T)$ of (45) with c=0, i.e., the Green function of K1. This martingale construction for solutions of Black-Scholes type pdes (45) is given in [5][1] using unnecessarily complicated notation, and without the explanation of the connection of the Black-Scholes pde with K1, K2, and the Chapman-Kolmogorov eqn.

Second, although 'memory' (or 'aftereffect') is never mentioned in the text [4], according to Friedman's definition processes with memory should labeled as 'Markov Processes' so long as the Chapman-Kolmogorov eqn. is satisfied. His definition of a Markov process (pg. 18, ref. [4]), stated here in terms of transition densities, is that (1) there exists a (Borel) measurable transition density $p(x,t:y,s) \geq 0$, (2) that $p(x,t:y,s)$ is a (probability) measure, and (3) that $p(x,t:y,s)$ satisfies the Chapman-Kolmogorov eqn. By this definition the Shimizu-Yamada model is Markovian. However, this classification contradicts the standard definition of 'Markov' as a process 'without aftereffect' [1,2], without history dependence [3,15]. A Markov process is typically defined as a process whereby the time evolution of the transition density $p(x,t:y,s)$ is fixed by specifying exactly one earlier state (y,s), s<t. For experts like Feller [16] and Doob [17] as well, the Chapman-Kolmogorov eqn. is a necessary but insufficient condition for a Markov process. The strength of Friedman's text is that it teaches us classes of diffusive nonMarkovian systems that satisfy that condition. Feller's example of a nonMarkov process satisfying the Chapman-Kolmogorov eqn. is discrete [16].

---

[1] In [5], eqns. (15.25) and (15.27) are inconsistent with each other, (15.25) cannot be obtained from (15.27) by a shift of coordinate origin because the x-dependent drift and diffusion coefficients break translation invariance. A careful treatment of solving elliptic and parabolic pdes by 'running a Brownian motion' is provided by Friedman [4].

**Acknowledgement**

JMC thanks Enrico Scalas, Harry Thomas, Gemunu Gunaratne, and Kevin Bassler for stimulating discussions via email about various points.

**References**

1. B. V. Gnedenko, *The Theory of Probability*, tr. by B.D. Seckler (Chelsea, N.Y., 1967).

2. R.L. Stratonovich. *Topics in the Theory of Random Noise*, tr. By R. A. Silverman (Gordon & Breach: N.Y 1963).

3. L. Arnold, *Stochastic Differential Equations* (Krieger, Malabar, 1992).

4. A. Friedman, *Stochastic Differential Equations and Applications* (Academic, N.Y., 1975).

5. J.M. Steele, *Stochastic Calculus and Financial Applications* (Springer-Verlag, N.Y., 2000).

6. K.E. Bassler, G.H. Gunaratne, & J. L. McCauley, *Hurst Exponents, Markov Processes, and Nonlinear Diffusion Equations*, *Physica A* **369**: 343 (2006).

7. J. L. McCauley , G.H. Gunaratne, & K.E. Bassler, *Martingales, Detrending Data, and the Efficient Market Hypothesis*, submitted (2007).

8. P. Hänggi and H. Thomas, *Time Evolution, Correlations, and Linear Response of Non-Markov Processes*, Zeitschr. Für Physik **B26**: 85 (1977).

9. J. L. McCauley, *Markov vs. nonMarkovian processes: A comment on the paper 'Stochastic feedback, nonlinear families of Markov processes, and nonlinear Fokker-Planck equations'* by T.D. Frank, submitted (2007).

10. J. L. McCauley, G.H. Gunaratne, & K.E. Bassler, *Hurst Exponents, Markov Processes, and Fractional Brownian Motion*, Physica ***A*** (2007).

11. P. Hänggi, H. Thomas, H. Grabert, and P. Talkner, *Note on time Evolution of Non-Markov Processes*, J. Stat. Phys. **18**: 155 (1978).

12. T.D. Frank, *Stochastic feedback, nonlinear families of Markov processes, and nonlinear Fokker-Planck equations*, Physica ***A331***: 391 (2004).

13. J.L. McCauley, *Dynamics of Markets: Econophysics and Finance* (Cambridge, Cambridge, 2004).

14. J. L. McCauley, G.H. Gunaratne, & K.E. Bassler, *Martingale Option Pricing, Physica **A*** (2007).

15. M.C. Wang & G.E. Uhlenbeck in *Selected Papers on Noise and Stochastic Processes*, ed. N. Wax, Dover: N.Y., 1954.

16. W. Feller, *The Annals of Math. Statistics* 30, No. 4, 1252, 1959.

17. J. L. Snell, *A Conversation with Joe Doob*, http://www.dartmouth.edu/~chance/Doob/conversation.html; *Statistical Science* **12**, No. 4, 301, 1997.